\documentclass[%
reprint,
footinbib,
amsmath,amssymb,
aps,pre,
nobibnotes
]{revtex4-2}

\usepackage[english]{babel}
\usepackage[utf8x]{inputenc}
\usepackage[colorinlistoftodos]{todonotes}
\usepackage{enumerate}
\usepackage{graphicx}% Include figure files
\usepackage{dcolumn}% Align table columns on decimal point
\usepackage{bm}% bold math
\usepackage{bbold}

\usepackage{textcomp}
\usepackage[pagewise,columnwise]{lineno}
\begin{document}

\preprint{APS/123-QED}

%\title{Nonextensivity arises from spontaneous symmetry breaking of temperature fluctuations}
%\title{Quasithermal distributions}
\title{Complexity drives the symmetry breaking of temperature fluctuations}

\author{D. Rosales Herrera}
\affiliation{Facultad de Ciencias F\'isico Matem\'aticas, Benem\'erita Universidad Aut\'onoma de Puebla, Apartado Postal 165, 72000 Puebla, Pue., M\'exico}

\author{M. Y. Gallegos Ruiz}
\affiliation{Facultad de Ciencias F\'isico Matem\'aticas, Benem\'erita Universidad Aut\'onoma de Puebla, Apartado Postal 165, 72000 Puebla, Pue., M\'exico}

\author{A. Fern\'andez T\'ellez}
\affiliation{Facultad de Ciencias F\'isico Matem\'aticas, Benem\'erita Universidad Aut\'onoma de Puebla, Apartado Postal 165, 72000 Puebla, Pue., M\'exico}

\author{J. R. Alvarado García}
\email{j.ricardo.alvarado@cern.ch}
\affiliation{Facultad de Ciencias F\'isico Matem\'aticas, Benem\'erita Universidad Aut\'onoma de Puebla, Apartado Postal 165, 72000 Puebla, Pue., M\'exico}

\author{J. E. Ram\'irez}
\email{jhony.eredi.ramirez.cancino@cern.ch}
\affiliation{Facultad de Ciencias F\'isico Matem\'aticas, Benem\'erita Universidad Aut\'onoma de Puebla, Apartado Postal 165, 72000 Puebla, Pue., M\'exico}

\begin{abstract}
%The study and understanding of the properties of systems transitioning from thermal to nonthermal states is an open question in physics, not only in thermodynamics but also in QCD and complex systems. 
Frequently, nonextensive systems are described through $q$-exponential functions, where the $q$-parameter determines how far a system is from equilibrium. 
In this work, we study systems constrained to $q-1\ll1$ to describe systems near thermal equilibrium, and derive the probability density function and temperature fluctuations. 
To this end, we derive the series expansion of the $q$-exponential function, yielding a perturbative series that is a superposition of gamma distributions. 
On the other hand, the temperature fluctuations are expressed as a superposition of Dirac delta functions and their derivatives, meaning that the bulk of the system remains at a constant temperature, while point-temperature variations emerge, marking the system's departure from thermal equilibrium.
The statistics of the temperature fluctuations exhibit a transition from a symmetric to a nonsymmetric distribution as the nonextensive parameter increases.
These results provide evidence of the association between the emergence of nonextensivity and the symmetry breaking of temperature fluctuations.
%We derive the probability density function and temperature fluctuations of systems with a nonextensive parameter satisfying $q-1\ll 1$. 
%To this end, we deduce the series expansion of the q-exponential function, which is expressed as a perturbative series where the zeroth order corresponds to the thermal distribution and the higher orders are expressed as superpositions of gamma distributions. On the other hand, the temperature fluctuations are expressed as a superposition of the Dirac delta functions and their derivatives, meaning that the bulk of the system remains at a constant temperature, while in the system emerge point temperature variations, which mark the departure of the system from the thermal equilibrium.
%The statistics of the temperature fluctuations exhibit a transition from a symmetric to a nonsymmetric distribution as the nonextensive parameter increases.
%%ESTO NO%%%These results provide the first evidence of the association between the emergence of nonextensivity and the spontaneous symmetry breaking of temperature fluctuations.
%These results provide the first evidence of nonextensivity as an emergent phenomenon arising from the symmetry breaking of temperature fluctuations.
\end{abstract}

\maketitle

%\emph{Introduction.-}
\section{Introduction}
\label{sec:intro}
In statistical physics, the Tsallis $q$-exponential distribution
\begin{equation}
    f_q(x)=\mathcal{N}_q [1-(1-q)\lambda x]^{1/(1-q)}
    \label{eq:qexp}
\end{equation}
was introduced as an alternative framework for physical phenomena that cannot be described by the statistical mechanics and thermodynamics of systems in equilibrium \cite{tsallis_possible_1988,Tsallis:2009zex}.
In Eq.~\eqref{eq:qexp}, $\lambda$ is a scale parameter, $q$ is a dimensionless parameter that quantifies the nonextensivity degree of the system, and $\mathcal{N}_q$ is the normalization constant.
Usually, the systems described through \eqref{eq:qexp} exhibit distinctive characteristics, such as long-range correlations and multifractality, and are described by heavy-tailed distributions.
The $q$-exponential distribution \eqref{eq:qexp} deviates from the conventional statistical mechanics characterized by Boltzmann’s exponential factor, which is recovered in the limit $q\to1$.
Applications include studies of turbulence in fluids~\cite{Beck2001turb,Arimitsu2002}, Lévy anomalous diffusion~\cite{Plastino1995}, complex networks~\cite{Abe2003internet}, statistics of cosmic rays~\cite{Tsallis2003cosmic}, the description of charged particle production in ultrarelativistic collisions~\cite{Cleymans2012,Bhattacharyya:2020sjd,Parvan2020,BHATTACHARYYA2022127836,Pajares:2022uts,Herrera:2024tyq}, econometry~\cite{Anteneodo2002,Borland2002}, and many others~\cite{TSALLIS200389,Picoli2009,Brito2016,Bhattacharyya:2017cdk}. 
In summary, the $q$-exponential distribution has been helpful in studying complex systems with high accuracy.

Interestingly, the $q$-exponential distribution has the following asymptotic limits
\begin{equation}
    f_q(x) \propto \begin{cases}
			\exp(-x/T) & \text{for $x\to0$}\\
            x^{1/(1-q)} & \text{for $x\to \infty$}
		 \end{cases},
\end{equation}
where $T=1/\lambda$ is usually denoted as the temperature of the system, since the $q$-exponential resembles the Boltzmann distribution at low values of the random variable.
In this way, we can introduce a formal definition of the temperature for a probability density function ($P(x)$) describing the distribution of a random variable $x$ if the following limit exists:
\begin{equation}
T=
-\lim_{x\to0}\left[
\frac{d}{dx}
\ln P(x)
\right]^{-1}.
\label{eq:Teff_def}
\end{equation}
Otherwise, alternative definitions should be used. 
For instance, following the definition of temperature in thermodynamics, it can be estimated as the derivative of the \emph{internal energy} (random variable) with respect to entropy (which can be computed as the Shannon entropy).
However, in this paper, we work with the temperature definition in \eqref{eq:Teff_def} since quasithermal distributions behave as exponential distributions in the limit $x\to0$.

%\adch{COMENTAR SOBRE LOS LÍMITES ASINTÓTICOS DE LA Q-EXPONENTIAL, INTRODUCIR LA TEMPERATURA COMO EL SLOPE DE LA DISTRIBUCIÓN?}

%\adch{MAS INTRODUCCIÓN}

%\adch{SCOPE}
In this work, we aim to study systems described by a $q$-exponential distribution, with $q-1\ll 1$, which are called quasithermal systems.
%.
%We called such distributions quasithermal distributions.
To this end, we compute the series expansion of the $q$-exponential distribution \eqref{eq:qexp} for $q\to1^+$, finding iterative formulas for determining the series expansion at arbitrary order.
Additionally, we determine the temperature and its fluctuations for such systems, which are computed as the slope of the distribution in the limit $x\to0$ and using the superstatistics framework. Then, we analyze different data sets within the quasithermal distributions.
The main result presented in this work is the description of the nonextensivity of the systems as an emergent phenomenon related to the symmetry breaking of temperature fluctuations.

The rest of the paper is organized as follows.
In Sec.~\ref{sec:quasithermals}, we derive the perturbative series expansion of the $q$-exponential distribution, called quasithermal distributions.
In Sec.~\ref{sec:stats}, we compute the statistics of the quasithermal distributions. 
In Sec.~\ref{sec:data}, we analyze diverse datasets with thermal, quasithermal, and $q$-exponential distributions. 
As goodness of fit metric, we estimate the coefficient of determination of the linearized data and models.  
We study the thermostatic properties of the quasithermal distributions in Sec.~\ref{sec:thermo}. 
In Sec.~\ref{sec:temperature_fluct}, we derive the temperature fluctuations associated with the quasithermal distributions. 
We discuss the applicability of quasithermal distributions to statistical models in Sec.~\ref{sec:applications}.
Finally, we wrap up our work with our concluding remarks in Sec.~\ref{sec:conclusions}.

\section{Perturbative series expansion of $q$-exponential distributions: Quasithermal distributions}
\label{sec:quasithermals}
%\emph{Perturbative series expansion of $q$-exponential distributions: Quasithermal distributions.-}

Let us start the derivation of the quasithermal distributions by setting $q=1+\epsilon$, where $0<\epsilon \ll 1$.
Considering the Taylor series expansion $\ln(1+z)=\sum_{n=1}^\infty (-1)^{n+1}z^n/n$ for $|z|<1$, we introduce the generating function $g(u,\epsilon)$ given by
\begin{equation}
    \ln g(u, \epsilon)=u-\frac{1}{\epsilon}\ln \left(1+\epsilon u  \right)=\sum_{n=2}^\infty (-1)^n \frac{u^n}{n}\epsilon^{n-1},
    \label{eq:lng}
\end{equation}
where $u=\lambda x$ has been introduced for the sake of notation.
Note that \eqref{eq:lng} is only valid if $u<1/\epsilon$. 
We found that the generating function can be rewritten as a product of exponential functions after exponentiation to both sides of Eq.~\eqref{eq:lng}.
Then, by considering the Taylor series expansion of each exponential factor, we found that
\begin{equation}
    g(u,\epsilon)=\exp(u)(1+\epsilon u)^{-1/\epsilon}=\sum_{n=0}^\infty \epsilon^n \mathbb{P}_n(u),
\end{equation}
with $\mathbb{P}_n(u)$ being a polynomial of degree $2n$.
It is straightforward to show that the polynomials satisfy $\mathbb{P}_0(u)=1$ and $\mathbb{P}_n(0)=0$ for $n\geq1$.
The latter relation acts as an initial condition to fully determine each polynomial $\mathbb{P}_n$, as we further discuss below.
Additionally, the generating function satisfies the relation $(1+\epsilon u)\partial_u g(u,\epsilon) = \epsilon ug(u,\epsilon)$. By substituting the series expansion of the generating function, we derive a recursive differential equation for determining the polynomials $\mathbb{P}_n$ for $n>0$, which is given by
\begin{equation}
    \mathbb{P}'_n(u)=u\left( \mathbb{P}_{n-1}(u)- \mathbb{P}'_{n-1}(u)  \right).
    \label{eq:recursive}
\end{equation}
The prime notation in Eq.~\eqref{eq:recursive} denotes the derivative of $\mathbb{P}_n$ with respect to $u$.
The polynomials can be written explicitly as
\begin{equation}
    \mathbb{P}_n(u)=\sum_{k=0}^{2n} \alpha_{n,k}u^k.
    \label{eq:genP}
\end{equation}
Substituting \eqref{eq:genP} in \eqref{eq:recursive} gives the recursive relationship between the coefficients of the polynomials $\mathbb{P}_n$:
\begin{equation}
   k \alpha_{n,k}=\alpha_{n-1, k-2}-(k-1)\alpha_{n-1, k-1}.
\end{equation}
Moreover, it is easy to prove that $\alpha_{n, k}=0$ for $k<n+1$ and $n>0$ \footnote{This claim can be proved by mathematical induction. 
\emph{Proof}. 
The statement is valid for $n=1$ since $\alpha_{11}=0$. 
Now, let us assume that $\alpha_{n1}=\alpha_{n2}=\dots=\alpha_{nn}=0$ for $\mathbb{P}_n(u)$. This implies that the right hand side of Eq.~\eqref{eq:recursive} is a polynomial of degree $2n+1$ and the term with the lowest degree is $(n+1)\alpha_{nn+1}u^{n}$. Therefore, after solving Eq.~\eqref{eq:recursive}, it is found that $\mathbb{P}_{n+1}(u)$ is a polynomial of degree $2n+2$ and the degree of the lowest exponent term is $n+2$.}.
The first polynomials are
\begin{align}
\mathbb{P}_0(u) &= 1, \nonumber
\\
\mathbb{P}_1(u) &= \frac{u^2}{2}, \nonumber
\\
\mathbb{P}_2(u) &= \frac{u^4}{8}-\frac{u^3}{3}, \nonumber
\\
\mathbb{P}_3(u) &= \frac{u^6}{48}-\frac{u^5}{6}+\frac{u^4}{4}, \nonumber
\\
\mathbb{P}_4(u) &= \frac{u^8}{384}-\frac{u^7}{24}+\frac{13}{72}u^6-\frac{u^5}{5}. \nonumber
%\label{eq:first_polys}
\end{align}
Interestingly, the polynomials $\mathbb{P}_n$ satisfy the relation $\int_0^\infty \exp(-u) \mathbb{P}_n(u) du=1$ for $n\geq0$, indicating that the function $\exp(-u)\mathbb{P}_n(u)$ constitutes by itself a well defined probability density function.
Then, the following relation for $\alpha_{n,k}$ is also satisfied
$\sum_{k=n+1}^{2n} \alpha_{n,k}k!=1.$

Finally, we introduce the quasithermal distribution at arbitrary order $m$ in $\epsilon$ as the following perturbative series 
\begin{align}
    \mathcal{Q}(x, \lambda,\epsilon, m)&=\lambda \exp(-u) \nonumber\\
    &+\lambda \exp(-u)\sum_{n=1}^m \epsilon^n \left( \mathbb{P}_n(u)-\mathbb{P}_{n-1}(u)  \right),
    \label{eq:quasiThermal}
\end{align}
where $u=\lambda x$.
Note that the quasithermal distribution in \eqref{eq:quasiThermal} also considers the normalization constant from the series expansion of the $q$-exponential distribution up to arbitrary order and it is well-normalized by integrating over the random variable $x$ and running the integral from zero to infinity.
%\chck{Notar que como la quasithermal está bien normalizada, eso significa que es una pdf bien definida, y se garantiza que los datos son reproducidos siempre que $\lambda x<1/\epsilon$. En caso contrario, si $\lambda x>1/\epsilon$, la distribución quasitérmica puede desviarse sustancialmente, sobre todo si los datos presentan una cola en ley de potencias muy pronunciada.}

It is possible to prove that quasithermal distributions~\eqref{eq:quasiThermal} take only positive values for $\lambda x<1/\epsilon$ at arbitrary order.
This property is generally not true for the entire support of the random variable, but it holds for $\epsilon<0.17$ up to $m=7$.
However, if a data set requires higher order quasithermal distributions, we strongly recommend fitting the $q$-exponential function.
On the other hand, for large $x$ values, Equation~\eqref{eq:quasiThermal} may deviate significantly from the data if they exhibit a pronounced power-law tail.

\begin{figure}
    \centering
    \includegraphics{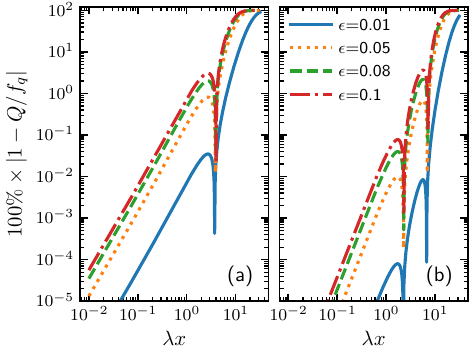}
\caption{
Absolute percentage difference between the quasithermal and the $q$-exponential distributions as a function of $\lambda x$, for (a) $m=1$ and (b) $m=2$. The distinct lines represent different values of the nonextensivity parameter: $\epsilon=0.01$ (solid), $\epsilon=0.05$ (dashed), $\epsilon=0.08$ (dash-dotted), and $\epsilon=0.1$ (dotted).
}
    \label{fig:dif}
\end{figure}

\section{Statistics of quasithermal distributions}
\label{sec:stats}
%\emph{Statistics of quasithermal distributions.-}

It is worth mentioning that the perturbative contribution of quasithermal distributions is expressed as a sum of terms of the form $u^k\exp(-u)$, which is the functional part of the gamma distribution. Thus, the quasithermal distribution can be rewritten in the following way
\begin{align}
    \mathcal{Q}(x, \lambda,\epsilon, m)&=\lambda \exp(-u)+\lambda \epsilon \left( \Gamma(u, 3)-\exp(-u)  \right) \nonumber\\
    &+\lambda \sum_{n=2}^m \epsilon^n \sum_{k=n}^{2n} \left( \alpha_{n, k}-\alpha_{n-1,k}  \right) k! \Gamma(u, k+1),
    \label{eq:quasiTG}
\end{align}
with $\Gamma(u, \alpha)=u^{\alpha-1}\exp(-u)/\Gamma(\alpha)$ being the gamma distribution.
We must recall that $\alpha_{n,k}\neq0$ if $n<k\leq2n$.
Equation~\eqref{eq:quasiTG} facilitates the interpretation of quasithermal distributions, which are given as a perturbative series around the thermal description. 
The presence of the gamma distribution terms raises the probability of observing large values of the random variable, but the tail of the quasithermal distributions remains dominated by the exponential behavior.
In Fig.~\ref{fig:dif}, we plot the absolute percentage difference between the quasithermal distributions up to first and second order and the $q$-exponential function for low values of $\epsilon$, finding a good agreement between them.

\begin{figure}
    \centering
    \includegraphics{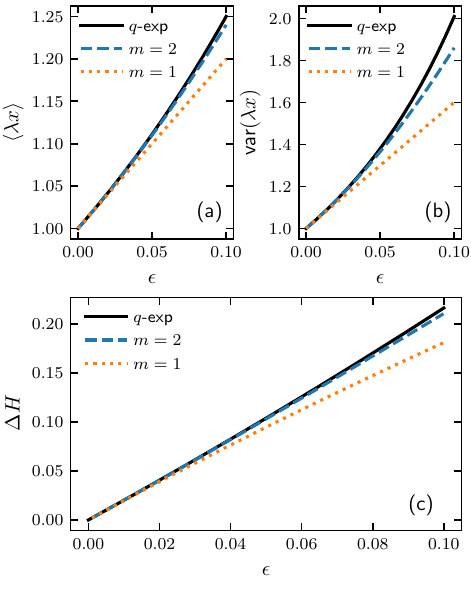}
    \caption{
(a) Mean value, (b) variance, and (c) entropy production of the quasithermal distributions for $m=1$ (dotted lines), $m=2$ (dashed lines), and the $q$-exponential function (solid lines) as a function of $\epsilon$. 
}
    \label{fig:statistics}
\end{figure}

Another advantage of Eq.~\eqref{eq:quasiTG} is that some relevant statistical properties of the quasithermal distributions can be written as a linear combination of the gamma distribution statistics. 
For instance, the characteristic function $\varphi(t)= \langle \exp(itx)  \rangle$ is
\begin{align}
    \varphi(t)&=\left(1-i\frac{t}{\lambda} \right)^{-1}+ \epsilon \left( \left(1-i \frac{t}{\lambda} \right)^{-3}-\left(1-i\frac{t}{\lambda} \right)^{-1}  \right) \nonumber\\
    &+\sum_{n=2}^m \epsilon^n \sum_{k=n}^{2n} \left( \alpha_{n, k}-\alpha_{n-1,k}  \right) k! \left(1-i\frac{t}{\lambda} \right)^{-(k+1)},
    \label{eq:CF}
\end{align}
from which the moment and cumulant generating functions can be computed as $M(t)=\varphi(-it)$ and $K(t)=\ln(\varphi(-it))$, respectively.
Then, the average and variance of the random variable $x$ are
\begin{align}
    \langle x \rangle &= M'(0)= \frac{1}{\lambda} \biggl(1+2\epsilon \nonumber \\ 
    &+ \sum_{n=2}^m \epsilon^n \sum_{k=n}^{2n} \left( \alpha_{n, k}-\alpha_{n-1,k}  \right) (k+1)! \biggl),\\
    \text{var}(x) &= K''(0)\approx\frac{1}{\lambda^2}+\frac{6}{\lambda^2}\epsilon \nonumber \\
    & +\frac{1}{\lambda^2}\sum_{n=2}^m \epsilon^n \sum_{k=n}^{2n} (\alpha_{n, k}-\alpha_{n-1, k})(k+1)! k \nonumber\\
    &- \frac{1}{\lambda^2} \sum_{j_0+\dots+j_m=2} \binom{2}{j_0, \dots, j_m} \epsilon^{\sum_{l=1}^{m} l j_l} \nonumber \\
    & \times \mathbb{1}_{\sum_{l=1}^{m} l j_l=0,\dots,m} \Sigma_1^{j_1}\times \cdots \times \Sigma_m^{j_m}, 
\end{align}
where $\mathbb{1}_{x=2,\dots,m}$ denotes the indicator function, which takes the value 1 if $x=0,\dots,m$ and zero otherwise.
The indicator function is used to express the variance as a perturbative series up to the $m$-th order in $\epsilon$.
The symbol $\Sigma_n$ denotes the factor multiplying the terms of the form $\epsilon^n/\lambda$, given by $\Sigma_1=2$ and
\begin{equation}
    \Sigma_n=\sum_{k=n}^{2n}(\alpha_{n, k}-\alpha_{n-1, k})(k+1)!.
\end{equation}
In Figs.~\ref{fig:statistics} (a) and (b), we plot $\langle \lambda x \rangle$ and $\text{var}(\lambda x)$, respectively.
In both cases, the second order series expansion is a good estimation of the average and variance for reproducing statistics of the $q$-exponential distribution in the case of $q$ values slightly above one (small values of $\epsilon$).
Here, we would like to comment that the convergence of the first and second moments of the $q$-exponential distribution is bounded by $q<3/2$ ($\epsilon<1/2$) and $q<4/3$ ($\epsilon<1/3$), respectively~\cite{Tsallis:1998ws,Herrera:2024zjy}. 
So, the quasithermal distribution up to the second order reproduces in good agreement for a considerable range of $\epsilon$ the corresponding statistics of the $q$-exponential distribution.

\section{Data analysis}
\label{sec:data}

\begin{figure*}
    \centering
    \includegraphics{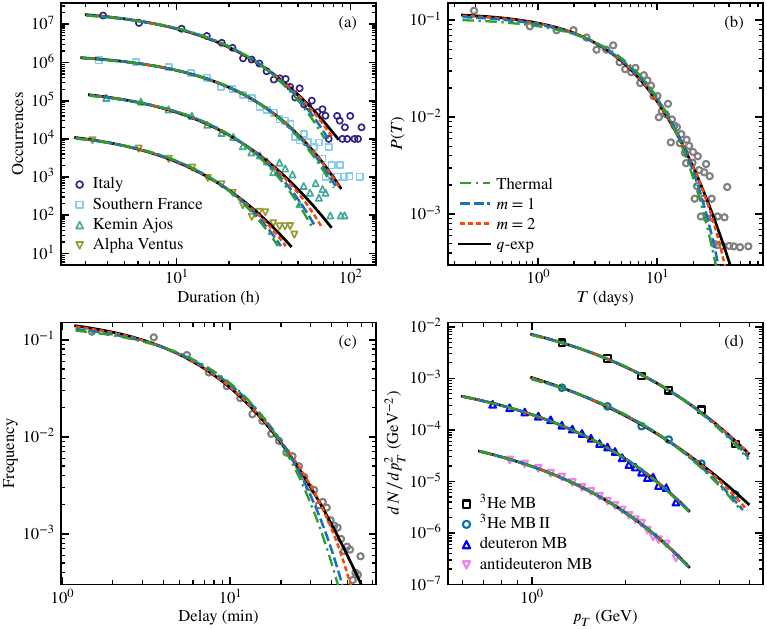}
    \caption{
Fit of the $q$-exponential and the quasithermal distributions to the analyzed data sets:
(a) Duration of low-wind persistence periods at four European locations~\cite{Weber2019}.
(b) Interevent time distribution of the declustered Greek seismicity catalog~\cite{Antonopoulos2014}.
(c) Train delay distribution on the Reading-London Paddington route~\cite{Briggs2007}.
(d) Transverse momentum spectra of light ions produced in pp collisions at different center of mass energies reported by the ALICE-LHC Collaboration: (anti)deuterons at $\sqrt{s}=7$~TeV~\cite{ALICE:2017xrp} and (anti)$^3$He at $\sqrt{s}=13$~TeV~\cite{ALICE:2021mfm}.
In all panels, the solid line corresponds to the $q$-exponential distribution, the dashed and dotted lines correspond to the quasithermal fits up to the first ($m=1$) and second ($m=2$) order, respectively. 
Thermal fits are plotted as dash-dotted lines.
Some data sets were scaled to improve visualization.
}
    \label{fig:fits}
\end{figure*}

\begin{figure}
    \centering
    \includegraphics{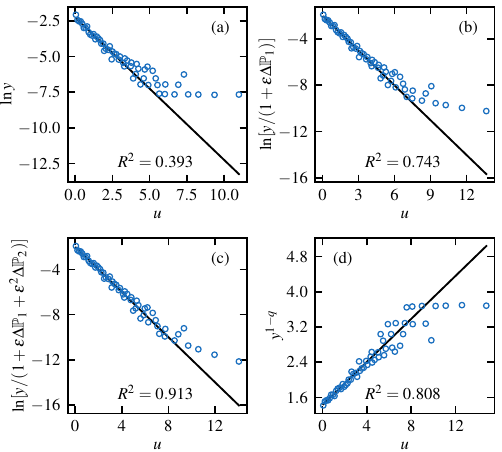}
    \caption{
Data linearization of the Greek seismicity interevent time for the models: 
(a) the thermal distribution, the quasithermal distribution at (b) first and (c) second truncation order, and (d) the $q$-exponential distribution.
In all panels, the solid line corresponds to the linearized models.
In the ``y''-axis labels of panels (b) and (c), $\Delta \mathbb{P}_n = \mathbb{P}_n - \mathbb{P}_{n-1}$.
    }
    \label{fig:lsample}
\end{figure}

\begin{figure}
    \centering
    \includegraphics{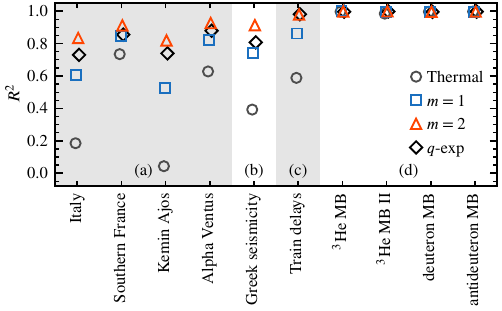}
    \caption{
Coefficient of determination ($R^2$) for each fitted model (markers) for all the analyzed data sets: (a) wind power persistence, (b) Greek seismicity, (c) train delays, and (d) light-nuclei production. 
}
    \label{fig:Rvals}
\end{figure}

We exemplify the applicability of quasithermal distributions by analyzing data sets from four different complex systems: wind power persistence~\cite{Weber2019}, seismic activity~\cite{Antonopoulos2014}, train delays~\cite{Briggs2007}, and light-nuclei production in ultrarelativistic collisions~\cite{ALICE:2017xrp,ALICE:2021mfm}.
In particular, we fit the quasithermal distribution up to first and second order to the data.
The method applied consists of an exhaustive search, and the best fit parameters are chosen as those minimizing the $\chi^2$. 
In all cases, the quasithermal distributions provide a good description of all considered data sets, as shown in Fig.~\ref{fig:fits}.
The inferred values of $\epsilon$ are less than 0.17, suggesting that the systems studied can be cataloged as quasithermal.
The latter means that the distributions related to those systems enhance their tail compared to the exponential distribution, but their nonextensivity is mild. 

We compute the coefficient of determination ($R^2$) of the linearized data for each model to assess the goodness of fit.
The linearization procedure is the following.
In the case of the thermal distribution, we take the logarithm of the frequencies.
For the quasithermal distributions, we first divide the frequencies by $1+\epsilon(\mathbb{P}_2-\mathbb{P}_1)$ and $1+\epsilon(\mathbb{P}_2-\mathbb{P}_1)+\epsilon^2(\mathbb{P}_3-\mathbb{P}_2)$ for $m=1$ and $m=2$, respectively.
For the case of the $q$-exponential, we take the frequencies to the power $1-q$.
In Fig.~\ref{fig:lsample}, we show the transformations of the seismic activity data for all models discussed in this paper.
After linearizing the data, we computed $R^2$ in the standard way using the corresponding linear form of each model. Figure~\ref{fig:Rvals} summarizes our $R^2$ estimates, showing that in all cases, the quasithermal distributions up to second order are the best fit across the models.

We would like to bring attention to the particular case of high energy physics data. The description of the transverse momentum spectrum of the light ion production in the form of quasithermal distributions suggests that the medium where such particles are created is almost thermal. Therefore, the light ion production is better described by the nonextensive charged particle production approach \cite{Herrera:2024tyq,RosalesHerrera:2026fmg} rather than the thermal model.

\section{Thermodynamics of quasithermal distributions}
\label{sec:thermo}
%\emph{Thermodynamics of quasithermal distributions.-}

Now, let us discuss some thermodynamic properties of the quasithermal distributions.
We recall that the temperature can be computed using Eq.~\eqref{eq:Teff_def}, resulting in $T=1/\lambda$ for the quasithermal distributions, which is expected since the exponential factor dominates over the polynomial sum at low $x$, as discussed above.
%Regarding the statistics of the quasithermal distributions, the $\ell$-th moment is also expressed as a perturbative series, given by
%\begin{equation}
%    \langle x^\ell \rangle=\frac{\ell!}{\lambda^\ell}+\frac{1}{\lambda^\ell}\sum_{n=1}^m \epsilon^n \left( I_{n\ell} - I_{n-1, \ell} \right),
%\end{equation}
%with
%\begin{equation}
%    I_{n\ell}=\int_{0}^\infty u^\ell \exp(-u) \mathbb{P}_n(u) du=\sum_{k=n+1}^{2n} \alpha_{n,k}(k+\ell)!
%\end{equation}
%for $n>0$. In the case of $n=0$, we have $I_{0\ell}=\ell!$.
%Thus,
%\begin{equation}
%    I_{n\ell}-I_{n,\ell-1}=\sum_{k=n+1}^{2n}\alpha_{n,k}(k+\ell-1)!(k+\ell-1).
%\end{equation}
%Thus, the expectation of $x$ is
%\begin{equation}
%    \langle x \rangle =T \frac{1 +\sum_{n=1}^m \sum_{k=n+1}^{2n} \epsilon^n \alpha_{n,k} (k+1)!}{1+\sum_{n=1}^m \sum_{k=n+1}^{2n} \epsilon^n \alpha_{n,k} k!}
%\end{equation}

On the other hand, for the computation of the Shannon entropy ($H=-\langle \ln \mathcal{Q} \rangle$), it is more convenient to use the quasithermal distributions given by Eq.~\eqref{eq:quasiThermal} because an exponential factor can be factorized. In consequence, it is possible to identify the contribution of the thermal description to the entropy. Thus, we obtain
\begin{align}
    H&=H_T + \langle \lambda x \rangle -1 \nonumber \\
    &- \left\langle \ln \left( 1+\sum_{n=1}^{m} \epsilon^n \left( \mathbb{P}_n(u) -\mathbb{P}_{n-1}(u)   \right) \right)  \right\rangle.
    \label{eq:H}
\end{align}
%We must recall that $\alpha_{n,k}=0$ if $k>2n$ and $k<n+1$.
In Eq.~\eqref{eq:H}, $H_T=1-\ln \lambda$ denotes the entropy of a thermal system, so the difference $\Delta H=H-H_T$ can be interpreted as the entropy production due to the perturbation of the system from the purely thermal behavior.
Note that the second and third terms in \eqref{eq:H} do not explicitly depend on temperature and only increase if $\epsilon$ does, as depicted in Fig.~\ref{fig:statistics} (c).
The latter means that the shape of the quasithermal distributions differs from the purely exponential distribution. 
In fact, the tail of $\mathcal{Q}(x, \lambda, \epsilon, m)$ rises, leading to an increase in the probability of observing rare events, which is in agreement with the heavy tail of the $q$-exponential distribution.
However, the heat capacity $C=-\lambda dH/d\lambda=1$ unless a particular process results in an increment of $\epsilon$, that is, $\epsilon$ is a function of the temperature.
In any case, $C>0$, indicating that quasithermal systems are thermodynamically stable.
For instance, it is observed that the production of charged particles in ultrarelativistic pp collisions raises their nonextensivity degree as the collisions become more energetic, which is consistent with the increment of the observation of high transverse momentum particles as the center of mass energy grows~\cite{Herrera:2024zjy}.

\section{Temperature fluctuations}
\label{sec:temperature_fluct}
%\emph{Temperature fluctuations.-}

It is possible to derive the temperature fluctuations throughout the system using the superstatistics approach \cite{Wilk:1999dr, Beck2003}. 
The main idea is to write the probability density function ($P(x)$) in the form of a Laplace transform, i. e., $P(x)=\int_0^\infty \exp(-\beta x) \mathcal{T}(\beta)$, where $\mathcal{T}(\beta)$ is the function describing the temperature fluctuations.
In particular, the convolution of the Gamma distribution with the thermal distribution gives the $q$-exponential distribution.
For the case of the quasithermal distributions, we recall that the probability density function is written as a linear combination of terms of the form $u^k \exp(-u)$ with $k$ being a positive integer.
Then, the unique way to express these terms as a Laplace transform is through the derivatives of the Dirac delta function, i. e., $\lambda e^{-\lambda x}(\lambda x)^k=\int_0^\infty
    \lambda^{k+1}\delta^{(k)}(\beta-\lambda) e^{-\beta x} d\beta$. 
Thus, we obtain the temperature fluctuations of the quasithermal distribution up to arbitrary order, given by  
\begin{align}
    \mathcal{T}_m(\beta)
    &=
    \lambda \delta(\beta-\lambda)
    -\lambda \epsilon \delta(\beta-\lambda)
    +\frac{1}{2}\lambda^3 \epsilon \delta^{(2)}(\beta-\lambda) \nonumber \\
    &+ \lambda \sum_{n=2}^m \epsilon^n \sum_{k=n}^{2n} \lambda^k (\alpha_{n, k}-\alpha_{n-1, k}) \delta^{(k)}  (\beta-\lambda),
    \label{eq:Tfluc}
\end{align}
which is not normalized and the computation of moments requires the proper normalization that considers the truncation of the perturbative series up to the desired order.

Before computing the statistics of the temperature fluctuations, we must recall that integrals involving derivatives of the Dirac delta function give
 \begin{equation*}
 \int_{-\infty}^{\infty} f(x) \delta^{(k)} (x) dx=(-1)^k f^{(k)} (0).      
 \end{equation*}
Therefore, the moments and central moments are given by
\begin{align}
\langle \beta^\ell \rangle=&\int_0^\infty \beta^\ell \widehat{\mathcal{T}}_m(\beta) d\beta, \label{eq:TFm}\\
\langle (\beta-\lambda)^\ell \rangle=&\int_0^\infty (\beta-\lambda)^\ell \widehat{\mathcal{T}}_m(\beta) d\beta, \label{eq:TFcm}
\end{align}
respectively. In Eqs.~\eqref{eq:TFm} and \eqref{eq:TFcm} $\widehat{\mathcal{T}}_m(\beta)$ denotes the normalized temperature fluctuations by $\lambda(1-\epsilon)$.
Note that
 \begin{equation*}
\int_0^\infty \beta^\ell \delta^{(n)}(\beta-\lambda)d\beta=\begin{cases}
			(-1)^n\frac{\ell !}{ (\ell -n)!} \lambda^{(\ell-n)}, & \text{if $\ell \geq n$}\\
            0, & \text{otherwise}
		 \end{cases}
 \end{equation*}
and
 \begin{equation*}
\int_0^\infty (\beta-\lambda)^\ell \delta^{(n)}(\beta-\lambda)d\beta=(-1)^nn! \delta_{\ell,n},
 \end{equation*}
with $\delta_{\ell,n}$ being the Kronecker delta.
Thus, the first moment and variance of the scaled random variable $\beta/\lambda$ are $\langle \beta/\lambda \rangle = 1$ and $\text{var}(\beta/\lambda)  = \epsilon$, respectively.
The third central moment is $\langle (\beta-\lambda)^3/\lambda^3 \rangle=2\epsilon^2$ for $m>1$ and 0 for $m=1$, from which we infer that the temperature fluctuations~\eqref{eq:Tfluc} have a positive skewness if $m>1$, but are symmetric for the first order approximation.

To interpret the temperature fluctuations~\eqref{eq:Tfluc}, we dip into the multipole expansion of the charge density in classical electrodynamics~\cite{Jackson1999}. 
In this context, a localized charge distribution at very large distances can be described as a superposition of point multipoles, represented as derivatives of the Dirac delta function. Thus, the first term in the temperature fluctuations is analogous to a monopole, representing the thermal term, while higher-order derivatives of the Dirac delta function introduce corrections due to the emergence of point temperature variations that break the thermal behavior.

\section{Applications of quasithermal distributions to statistical models}
\label{sec:applications}
%\emph{Applications of quasithermal distributions to statistical models.-}

Additionally, quasithermal distributions can be applied to statistical models where a particular observable is written as a joint bivariate probability distribution $O=M(v, w)$ \cite{Feller1971}. 
If the model admits fluctuations in $w$, then the observable should be computed as the marginal distribution, that is, $O(v)=\int_{\Omega_w}M(v, w)f(w)dw$.
If $w$ follows a quasithermal distribution, the probability density function for the observable $O$ is
\begin{align}
    O(v)&=\lambda \mathcal{L}_w\{M(v, w)\}
    +\lambda\sum_{n=1}^m \epsilon^n \sum_{k=n-1}^{2n}(\alpha_{n,k}-\alpha_{n-1,k}) \nonumber\\ 
    & \times (-\lambda)^k\frac{\partial^k}{\partial \lambda^k}\mathcal{L}_w\{M(v, w)\}, 
    \label{eq:OL}
\end{align}
where $\mathcal{L}_w\{M(v,w)\}$ denotes the Laplace transform of $M(v, w)$ performed only over the variable $w$.
It is straightforward to compute the characteristic function of $O$ in Eq.~\eqref{eq:OL}, yielding a similar expression to \eqref{eq:OL} where the function $\varphi_v(t,w)=\int_0^\infty e^{itv}M(v,w) dv$ takes the place of $M(v, w)$.
Thus, the $\ell$-moment is given by
\begin{align}
    \langle v^\ell \rangle &= \lambda \mathcal{L}_w \left \{ \left\langle v^\ell \right\rangle_v \right \}
    +\lambda\sum_{n=1}^m\epsilon^n \sum_{k=n-1}^{2n}(\alpha_{n,k}-\alpha_{n-1,k}) \nonumber \\
    &\times (-\lambda)^k \frac{\partial^k}{\partial \lambda^k}\mathcal{L}_w\left \{ \left\langle v^\ell \right\rangle_v \right \},
\end{align}
with $\left\langle v^\ell \right\rangle_v=\int_0^\infty v^\ell M(v, w) dv$.
This result implies that the statistics of models with quasithermal fluctuations can be computed from the Laplace transforms of the moments of the kernel $M(v, w)$ and their derivatives, which may simplify their estimation.
Examples of such models include the statistical nucleosynthesis model for describing the abundances of light elements in the early universe, for which the fluctuations are described by a $q$-exponential function with $q$ in the range 1.069-1.082~\cite{Hou:2017uap}.
Another example comes from ultrarelativistic collisions~\cite{becattiniheinz1997,becattinipassaleva2002}.
The production of light ions in high energy physics experiments is reported to be thermal~\cite{ALICE:2015wav}. 
Nevertheless, as we reported in this paper, the data are well described by the quasithermal distribution up to the second order. Therefore, the medium in which these particles are created is no longer thermal~\cite{ALICE:2025byl} and admits a description in terms of nonextensive particle production~\cite{Herrera:2024tyq,RosalesHerrera:2026fmg}.

\section{Conclusions}
\label{sec:conclusions}
%\emph{Conclusions.-}

In summary, in this work we derived the probability distribution describing systems near thermal equilibrium with a nonextensive parameter constrained to $q-1\ll 1$. 
This was done by deducing the series expansion of the $q$-exponential function around $q=1$, yielding a perturbative series in which the zeroth-order term coincides with the thermal distribution and higher orders are a superposition of gamma distributions.
In this way, we introduced the quasithermal distributions as the perturbative series up to arbitrary order, given by Eqs.~\eqref{eq:quasiThermal} and \eqref{eq:quasiTG}.
As we discussed above, the quasithermal distribution up to the second order can reproduce with high precision the behavior and statistics of the $q$-exponential distribution.
It was also shown that the quasithermal distribution with $m=2$ is a good fitting function in the cases of $q$ close to one (see Fig.~\ref{fig:fits}).

Another important result presented in this paper is the derivation of the temperature fluctuations of the quasithermal distributions, which is expressed as a superposition of the Dirac delta functions and their derivatives (see Eq.~\eqref{eq:Tfluc}).
It can be interpreted similarly to the multipole expansion of the charge density in classical electrodynamics. Therefore, quasithermal systems maintain a constant temperature in the bulk. However, as the nonextensive degree increases, point temperature variations begin to appear. These variations are of zero measure since the heat capacity of the system aligns with the thermal case. Nevertheless, quasithermal systems are out of equilibrium by definition.
Here, it is important to point out that the temperature fluctuations become broader as the nonextensive parameter grows, indicating that in quasithermal systems, the range of possible variations of the temperature becomes large, giving way to the emergence of more point temperature variations that eventually will start to form small regions across the system, resembling the interpretation of Wilk and W\l{}odarczyk of the temperature fluctuations associated with the $q$-exponential distribution.
Here, we must pay attention to the skewness, which is zero for $m=1$, but $m>1$.
This result means that temperature fluctuations are symmetric as the first order expansion dominates, whereas the fluctuations become asymmetric when the second order term significantly contributes to the perturbative expansion.
Overall, our results indicate that nonextensivity is an emergent phenomenon arising from the symmetry breaking of temperature fluctuations.

\acknowledgments
%\emph{Acknowledgments.-}
This work was funded by Secretaría de Ciencia, Humanidades, Tecnología e Innovación (SECIHTI-México) under the project CF-2019/2042, graduate fellowship grant number 1140160, and postdoctoral fellowship grant number 645654.

\emph{Data availability.-} There are no publicly available research data or software supporting this manuscript. Requests for further information or data should be sent to the authors.

\bibliography{ref}

@article{Weber2019,
  author  = {Weber, Juliane and Reyers, Mark and Beck, Christian
             and Timme, Marc and Pinto, Joaquim G.
             and Witthaut, Dirk and Sch{\"a}fer, Benjamin},
  title   = {Wind Power Persistence Characterized by Superstatistics},
  journal = {Sci. Rep.},
  volume  = {9},
  pages   = {19971},
  year    = {2019},
  doi     = {10.1038/s41598-019-56286-1}
}

@article{Antonopoulos2014,
  author  = {Antonopoulos, Chris G. and Michas, George
             and Vallianatos, Filippos and Bountis, Tassos},
  title   = {Evidence of $q$-exponential statistics in {Greek} seismicity},
  journal = {Physica A},
  volume  = {409},
  pages   = {71--77},
  year    = {2014},
  doi     = {10.1016/j.physa.2014.04.042}
}

@article{Briggs2007,
  author  = {Briggs, Keith and Beck, Christian},
  title   = {Modelling train delays with $q$-exponential functions},
  journal = {Physica A},
  volume  = {378},
  pages   = {498--504},
  year    = {2007},
  doi     = {10.1016/j.physa.2006.11.084}
}

@article{ALICE:2017xrp,
    author = "Acharya, Shreyasi and others",
    collaboration = "ALICE",
    title = "{Production of deuterons, tritons, $^{3}$He nuclei and their antinuclei in pp collisions at $\mathbf{\sqrt{{\textit s}}}$ = 0.9, 2.76 and 7 TeV}",
    doi = "10.1103/PhysRevC.97.024615",
    journal = "Phys. Rev. C",
    volume = "97",
    number = "2",
    pages = "024615",
    year = "2018"
}

@article{ALICE:2021mfm,
    author = "Acharya, Shreyasi and others",
    collaboration = "ALICE",
    title = "{Production of light (anti)nuclei in pp collisions at $ \sqrt{s} $ = 13 TeV}",
    doi = "10.1007/JHEP01(2022)106",
    journal = "JHEP",
    year = "2022",    
    volume = "01",
    pages = "106"
}

@article{Herrera:2024tyq,
    author = "Rosales Herrera, D. and Alvarado Garc\'\i{}a, J. R. and Fern\'andez T\'ellez, A. and Ram\'\i{}rez, J. E. and Pajares, C.",
    title = "{Nonextensivity and temperature fluctuations of the Higgs boson production}",
    doi = "10.1103/PhysRevC.110.015205",
    journal = "Phys. Rev. C",
    volume = "110",
    number = "1",
    pages = "015205",
    year = "2024"
}

@book{Tsallis:2009zex,
    author = "Tsallis, Constantino",
    title = "Introduction to Nonextensive Statistical Mechanics: Approaching a Complex World",
    doi = "10.1007/978-0-387-85359-8",
    isbn = "978-0-387-85358-1, 978-0-387-85359-8",
    publisher = "Springer",
    address = "New York",
    year = "2009"
}

@article{Wilk:1999dr,
    author = "Wilk, G. and W\l{}odarczyk, Z.",
    title = "{On the interpretation of nonextensive parameter q in Tsallis statistics and Levy distributions}",
    reportNumber = "SINS-PVIII-1999-9",
    doi = "10.1103/PhysRevLett.84.2770",
    journal = "Phys. Rev. Lett.",
    volume = "84",
    pages = "2770",
    year = "2000"
}

@article{RosalesHerrera:2026fmg,
    author = "Rosales Herrera, D. and Calder{\'o}n Mu{\~n}oz, J. C. and Alvarado Garc{\'\i}a, J. R. and Fern{\'a}ndez T{\'e}llez, A. and Ram{\'\i}rez, J. E.",
    title = "{Nonextensive Description of Charged-Particle Production in Ultrarelativistic Collisions}",
    doi = "10.3390/e28030298",
    journal = "Entropy",
    volume = "28",
    number = "3",
    pages = "298",
    year = "2026"
}

@book{Jackson1999,
  author    = {Jackson, John David},
  title     = {Classical Electrodynamics},
  edition   = {3rd},
  publisher = {John Wiley \& Sons},
  address   = {New York},
  year      = {1999},
  isbn      = {978-0-471-30932-1}
}

@article{Beck2003,
  author    = {Beck, Christian and Cohen, E. G. D.},
  title     = {Superstatistics},
  journal   = {Physica A},
  volume    = {322},
  pages     = {267--275},
  year      = {2003},
  doi       = {10.1016/S0378-4371(03)00019-0},
  publisher = {Elsevier}
}

@article{Hou:2017uap,
    author = "Hou, S. Q. and He, J. J. and Parikh, A. and Kahl, D. and Bertulani, C. A. and Kajino, T. and Mathews, G. J. and Zhao, G.",
    title = "{Non-extensive Statistics to the Cosmological Lithium Problem}",
    doi = "10.3847/1538-4357/834/2/165",
    journal = "Astrophys. J.",
    volume = "834",
    number = "2",
    pages = "165",
    year = "2017"
}

@article{Tsallis:1998ws,
    author = "Tsallis, C. and Mendes, R. S. and Plastino, A. R.",
    title = "{The Role of constraints within generalized nonextensive statistics}",
    doi = "10.1016/S0378-4371(98)00437-3",
    journal = "Physica A",
    volume = "261",
    pages = "534",
    year = "1998"
}

@article{Herrera:2024zjy,
    author = "Rosales Herrera, D. and Alvarado Garc\'{\i}a, J. R. and Fern\'andez T\'ellez, A. and Ram\'{\i}rez, J. E. and Pajares, C.",
    title = "{Entropy and heat capacity of the transverse momentum distribution for $pp$ collisions at RHIC and LHC energies}",
    doi = "10.1103/PhysRevC.109.034915",
    journal = "Phys. Rev. C",
    volume = "109",
    issue = "3",
    pages = "034915",
    year = "2024"
}

@article{ALICE:2015wav,
    author = "Adam, Jaroslav and others",
    collaboration = "ALICE",
    title = "{Production of light nuclei and anti-nuclei in pp and Pb-Pb collisions at energies available at the CERN Large Hadron Collider}",
    doi = "10.1103/PhysRevC.93.024917",
    journal = "Phys. Rev. C",
    volume = "93",
    number = "2",
    pages = "024917",
    year = "2016"
}

@article{ALICE:2025byl,
    author = "Acharya, S. and others",
    collaboration = "ALICE",
    title = "{Observation of deuteron and antideuteron formation from resonance-decay nucleons}",
    doi = "10.1038/s41586-025-09775-5",
    journal = "Nature",
    volume = "648",
    number = "8093",
    pages = "306--311",
    year = "2025"
}

@article{Pajares:2022uts,
    author = "Pajares, C. and Ram\'\i{}rez, J. E.",
    title = "{On the relation between the soft and hard parts of the transverse momentum distribution}",
    doi = "10.1140/epja/s10050-023-01170-w",
    journal = "Eur. Phys. J. A",
    volume = "59",
    number = "11",
    pages = "250",
    year = "2023"
}

@article{Picoli2009,
  author    = {Picoli, S. Jr. and Mendes, R. S. and Malacarne, 
               L. C. and Papa, R. P. T.},
  title     = {$q$-distributions in complex systems: a brief review},
  journal   = {Braz. J. Phys.},
  volume    = {39},
  number    = {2A},
  pages     = {468--474},
  year      = {2009},
  doi       = {10.1590/S0103-97332009000400023}
}

@article{TSALLIS200389,
title = {Nonextensive statistical mechanics and economics},
journal = {Physica A},
volume = {324},
number = {1},
pages = {89-100},
year = {2003},
issn = {0378-4371},
doi = {10.1016/S0378-4371(03)00042-6},
author = {Constantino Tsallis and Celia Anteneodo and Lisa Borland and Roberto Osorio}
}

@article{tsallis_possible_1988,
	title = {Possible generalization of {Boltzmann}-{Gibbs} statistics},
	volume = {52},
	issn = {1572-9613},
	url = {https://doi.org/10.1007/BF01016429},
	doi = {10.1007/BF01016429},
	number = {1},
	journal = {J. Stat. Phys.},
	author = {Tsallis, Constantino},
	month = jul,
	year = {1988},
	pages = {479--487},
}

@article{Beck2001turb,
  author  = {Beck, C. and Lewis, G. S. and Swinney, H. L.},
  title   = {Measuring nonextensitivity parameters in a turbulent
             {Couette-Taylor} flow},
  journal = {Phys. Rev. E},
  volume  = {63},
  pages   = {035303},
  year    = {2001},
  doi     = {10.1103/PhysRevE.63.035303}
}

@article{Arimitsu2002,
  author  = {Arimitsu, T. and Arimitsu, N.},
  title   = {Analysis of fully developed turbulence in terms of
             {Tsallis} statistics},
  journal = {Physica A},
  volume  = {305},
  pages   = {218--226},
  year    = {2002},
  doi     = {10.1016/S0378-4371(01)00665-3}
}

@article{Tsallis2003cosmic,
  author  = {Tsallis, C. and Anjos, J. C. and Borges, E. P.},
  title   = {Fluxes of cosmic rays: a delicately balanced
             stationary state},
  journal = {Phys. Lett. A},
  volume  = {310},
  pages   = {372--376},
  year    = {2003},
  doi     = {10.1016/S0375-9601(03)00377-3}
}

@article{Anteneodo2002,
  author  = {Anteneodo, C. and Tsallis, C. and Martinez, A. S.},
  title   = {Risk aversion in economic transactions},
  journal = {Europhys. Lett.},
  volume  = {59},
  pages   = {635--641},
  year    = {2002},
  doi     = {10.1209/epl/i2002-00172-5}
}

@article{Borland2002,
  author  = {Borland, L.},
  title   = {Option pricing formulas based on a non-{Gaussian}
             stock price model},
  journal = {Phys. Rev. Lett.},
  volume  = {89},
  pages   = {098701},
  year    = {2002},
  doi     = {10.1103/PhysRevLett.89.098701}
}

@article{Plastino1995,
  author  = {Plastino, A. R. and Plastino, A.},
  title   = {Non-extensive statistical mechanics and generalized
             {Fokker-Planck} equation},
  journal = {Physica A},
  volume  = {222},
  pages   = {347--354},
  year    = {1995},
  doi     = {10.1016/0378-4371(95)00211-1}
}

@article{Cleymans2012,
  author  = {Cleymans, J. and Worku, D.},
  title   = {The {Tsallis} distribution in proton-proton collisions
             at {LHC} energies at the {CERN} Large Hadron Collider},
  journal = {J. Phys. G},
  volume  = {39},
  pages   = {025006},
  year    = {2012},
  doi     = {10.1088/0954-3899/39/2/025006}
}

@article{Brito2016,
  author  = {Brito, S. G. A. and da Silva, L. R. and Tsallis, C.},
  title   = {Role of dimensionality in complex networks: 
             connection with nonextensive statistics},
  journal = {Sci. Rep.},
  volume  = {6},
  pages   = {27992},
  year    = {2016},
  doi     = {10.1038/srep27992}
}

@article{Abe2003internet,
  author  = {Abe, S. and Suzuki, N.},
  title   = {Itineration of the {Internet} over nonequilibrium 
             stationary states in {Tsallis} statistics},
  journal = {Phys. Rev. E},
  volume  = {67},
  pages   = {016106},
  year    = {2003},
  doi     = {10.1103/PhysRevE.67.016106}
}

@article{Parvan2020,
  author  = {Parvan, A. S. and Bhattacharyya, T.},
  title   = {Hadron transverse momentum distributions of the 
             {Tsallis} normalized and unnormalized statistics},
  journal = {Eur. Phys. J. A},
  volume  = {56},
  pages   = {72},
  year    = {2020},
  doi     = {10.1140/epja/s10050-020-00083-2}
}

@article{Bhattacharyya:2020sjd,
    author = "Bhattacharyya, Trambak and Parvan, Alexandru S.",
    title = "{Analytical results for the classical and quantum Tsallis hadron transverse momentum spectra: the zeroth order approximation and beyond}",
    doi = "10.1140/epja/s10050-021-00527-3",
    journal = "Eur. Phys. J. A",
    volume = "57",
    number = "6",
    pages = "206",
    year = "2021"
}

@article{Bhattacharyya:2017cdk,
    author = "Bhattacharyya, T. and Cleymans, J. and Mogliacci, S. and Parvan, A. S. and Sorin, A. S. and Teryaev, O. V.",
    title = "{Non-extensivity of the QCD p$_{T}$-spectra}",
    doi = "10.1140/epja/i2018-12647-6",
    journal = "Eur. Phys. J. A",
    volume = "54",
    number = "12",
    pages = "222",
    year = "2018"
}

@article{BHATTACHARYYA2022127836,
title = {A Tsallis-like effective exponential delay discounting model and its implications},
journal = {Physica A},
volume = {603},
pages = {127836},
year = {2022},
issn = {0378-4371},
doi = {10.1016/j.physa.2022.127836},
author = {Trambak Bhattacharyya and Shanu Shukla and Ranu Pandey},
}

@book{Feller1971,
  author    = {Feller, William},
  title     = {An Introduction to Probability Theory and Its Applications},
  volume    = {2},
  edition   = {2},
  publisher = {John Wiley \& Sons},
  address   = {New York},
  year      = {1971},
  isbn      = {978-0-471-25709-7},
}

@article{becattinipassaleva2002,
  author  = {F. Becattini and G. Passaleva},
  title   = {Statistical hadronization model and transverse momentum spectra of hadrons in high energy collisions},
  journal = {Eur. Phys. J. C},
  volume  = {23},
  pages   = {551--583},
  year    = {2002},
  doi     = {10.1007/s100520100869}
}

@article{becattiniheinz1997,
  author  = {F. Becattini and U. Heinz},
  title   = {Thermal hadron production in pp and p\={p} collisions},
  journal = {Z. Phys. C},
  volume  = {76},
  pages   = {269--286},
  year    = {1997},
  doi     = {10.1007/s002880050551}
}

\end{document}